\documentclass[a4paper,11pt]{article}
\usepackage{pos}
\usepackage{natbib}
\usepackage{subfig}
\graphicspath{ {figures/} }

\title{Scattering of dark pions in an Sp(4) gauge theory}

\author*[a]{Yannick Dengler}
\author[a]{Axel Maas}
\author[a,b]{Fabian Zierler}

\affiliation[a]{University of Graz, Universitätsplatz 5, 8010 Graz, Austria}
\affiliation[b]{Swansea University, Singleton Park, SA2 8PP Swansea, Wales, United Kingdom}

\emailAdd{yannick.dengler@uni-graz.at}
\emailAdd{fabian.zierler@swansea.ac.uk}
\emailAdd{axel.maas@uni-graz.at}

\abstract{
In this work we consider strongly interacting dark matter candidates as composite states of $N_f=2$ fermions charged under a dark $Sp(4)$ gauge group in the fundamental representation. We give expressions that allow the calculation of correlation functions of two pseudo-Nambu-Goldstone-bosons with lattice field theory and present first results on the scattering phase shift in the isospin-2 channel in the theory from first principles. We give a lower limit on the dark matter particle mass by comparing our results with astrophysical constraints on the cross-section.}

\FullConference{
The 40th International Symposium on Lattice Field Theory (Lattice 2023)\\
July 31st - August 4th, 2023\\
Fermi National Accelerator Laboratory
}

\begin{document}
\maketitle

\section{Introduction}
Dark matter is an attempt to describe a series of astrophysical and cosmological phenomena that cannot be explained in the scope of the Standard Model of particle physics (SM), the best known example being galactic rotation curves \cite{Rubin:1978kmz}. Although there is a lot of experimental and theoretical effort to identify dark matter, its precise nature is still elusive. 

Strongly-interacting models of dark matter have gained some interest as they might be able to resolve small-scale structure problems \cite{Tulin:2017ara}. Among them, the SIMP (\textbf{S}trongly \textbf{I}nteracting \textbf{M}assive \textbf{P}article) models provide dark matter as a thermal relic in a freeze-out process driven by a $3\to2$ semi-annihilation process. Such a process arises naturally for gauge theories with at least five pseudo-Nambu-Goldstone bosons (pNGBs) that we call (dark) pions arising from spontaneous chiral symmetry breaking via the Wess-Zumino-Witten term \cite{Hochberg:2014kqa}.

Symplectic gauge theories with two fundamental Dirac fermions provide a minimal realisation of the SIMP model, because the global symmetries are enlarged due to the pseudoreality of the fundamental representation \cite{Kulkarni:2022bvh}. Such theories furnish a $U(2N_f)$ flavour symmetry that is broken to $SU(2N_f)$ by the axial anomaly. Chiral symmetry breaking as well as explicit fermion masses break the symmetry further to $Sp(2N_f)$. The breaking gives rise to the pNGBs,  which we will call (dark) pions, in analogy to QCD. In such a theory $N_f$ fermions in a pseudoreal representation yield $2 N_f^2 - N_f - 1$ pNGBs. Thus, for $N_f=2$ the minimally required number of five pNGBs is obtained. This has been studied previously extensively on the lattice for the simplest case of an $SU(2)=Sp(2)$ gauge group. Here, we consider the next-simplest case, an $Sp(4)$ gauge group, motivated by the generally looser experimental constraints at larger $N_c$ \cite{Hochberg:2014kqa}.

One constraint on the properties of dark matter are the indirectly inferred limits on dark matter self-interactions at low center-of-mass energy from astrophysical observations \cite{Eckert:2022qia}. We address this in our model by calculating the scattering properties of the dark pions in the isospin-2 scalar channel. This channel is statistically the most likely for a random isospin distribution in cosmic matter. To access this information we determined the phase shifts in a L\"uscher-type analysis. A secondary use of our results is in the context of composite Higgs models \cite{Bennett:2017kga}.

\section{Lattice}

We employ lattice calculations using the HiRep code \cite{DelDebbio:2008zf}. The unimproved Wilson action is used for both the gauge fields and two dynamical fermions in the fundamental representation. We study fermion masses corresponding to values of $\frac{m_\pi}{m_\rho}$ between 0.65 and 0.90 at three distinct values for the inverse gauge coupling $\beta = 6.9, 7.05$ and $7.2$.

\subsection{Correlation functions}

Correlation functions are obtained by Wick-contracting operators. Using the definitions of the pion operators from \cite{Drach:2021uhl} we build the following isospin-0 and isospin-2 operators
\begin{align}
\label{Wick_Operators}
\begin{split}
    \mathcal{O}_{\pi\pi}^{I=2} &= \pi^+\pi^+ \\
    \mathcal{O}_{\pi\pi}^{I=0} &= \frac{1}{\sqrt{5}}\left(\pi^+\pi^-+\pi^-\pi^+-\pi^0\pi^0+\Pi_{ud}\Pi_{\Bar{u}\Bar{d}}+\Pi_{\Bar{u}\Bar{d}}\Pi_{ud}\right).
\end{split}
\end{align}
Note that we differentiate here between with $\pi$ and $\Pi$ the operators of magnetic isospin quantum numbers $\text{m}\le|1|$ and $\text{m}=|2|$, respectively. The latter can be interpreted as diquark states and are not present in a theory with a complex fermion representation of the gauge group. We omit space-time indices. The correlation functions are
\begin{align}
    C^I_{\pi\pi}(\tau) = \left\langle \mathcal{O}_{\pi\pi}^I(t=0)\mathcal{O}^{\dagger I}_{\pi\pi}(t=\tau) \right\rangle.
\end{align}
They can also be represented diagrammatically, as is done for example in \cite{RBC:2023xqv,Janowski:2019svg,Drach:2021uhl}. We adapt the notation from \cite{RBC:2023xqv} for the Wick contractions that resemble the "disconnected" (D), "cross" (C), "rectangle" (R) and "vacuum" (V) diagrams. In that notation they read,
\begin{align}
\label{Wick_contractions_02}
\begin{split}
    C_{\pi\pi}^{I=2} &= 2D-2C \\
    C_{\pi\pi}^{I=0} &= 2D+3C-10R+5V.
\end{split}
\end{align}
The isospin-2 channel coincides with QCD \cite{RBC:2023xqv}, because it does not get a contribution from mixing with the diquark operators. The isospin-0 channel however gets contributions from all pion operators including the diquark operators and therefore differs from QCD \cite{Drach:2021uhl}. This channel suffers from more numerical noise because it contains vacuum diagrams in which quarks propagate from and to the same time slice. For the expressions in the isospin-1 channel, we refer to \cite{Janowski:2019svg}. 

We use $Z_2 \times Z_2$ stochastic noise sources with spin-dilution \cite{Foley:2005ac} for the fermion sources. We remove constant contributions to the two-pion-correlator due to "around-the-world" effects by applying a numerical derivative \cite{Umeda:2007hy}. Afterwards, we use the corrfitter package to extract energy levels \cite{peter_lepage_2021_5733391}. We note that for our results we only use the one operator shown above for the isospin-2 channel. For improved systematics, one should include more operators and perform a variational analysis.

\subsection{Lüscher Analysis}
\label{Lüscher_section}

Interactions shift the finite volume energy levels. The Lüscher analysis is a tool to relate these energy shifts to infinite volume scattering properties. We employ here two particles with vanishing center-of-mass momentum only. The formalism is valid for energy levels between the elastic and the first inelastic threshold $( 2m_\pi < E < 4m_\pi$, in our case$)$. The finite volume energy levels can be translated to the lattice momentum $p^*$ via a dispersion relation accounting for the periodicity of the lattice,
\begin{align}
\begin{split}
    \cosh\left(\frac{E}{2}\right) = \cosh(m)+2\sin\left(\frac{p^*}{2}\right)^2, ~~~~~~~~~~~~~~
    q = p^*\frac{L}{2\pi}.
    \label{eq:dispersion_relation}
\end{split}
\end{align}
where we defined the generalized momentum $q$. The phase shift $\delta_0$ is calculated by the following formula which includes the transcendental Zeta function $\mathcal{Z}$. 
\begin{align}
\begin{split}
   \tan(\delta_0(q))=\frac{\pi^{\frac{3}{2}}q}{\mathcal{Z}^{\Vec{0}}_{00}(1,q^2)}
   \label{eq:tan_PS}
\end{split}
\end{align}
We refer to \cite{Rummukainen:1995vs,Jenny:2022atm} for details on the calculation of the Zeta function. Close to threshold, scattering follows a universal behavior described by a power series in $p^{*2}$. 
\begin{align}
\begin{split}
   |\Vec{p}|\cot(\delta_0(q)) = \frac{1}{a_0} + \mathcal{O}(p^{*2})
   \label{eq:p_cot_PS}
\end{split}
\end{align}
In particular, the leading constant term can be identified with the inverse scattering length $a_0^{-1}$ \cite{Meissner:2022cbi}.

\section{Results}
\begin{figure}
\begin{center}
    \includegraphics[width = 0.90\textwidth]{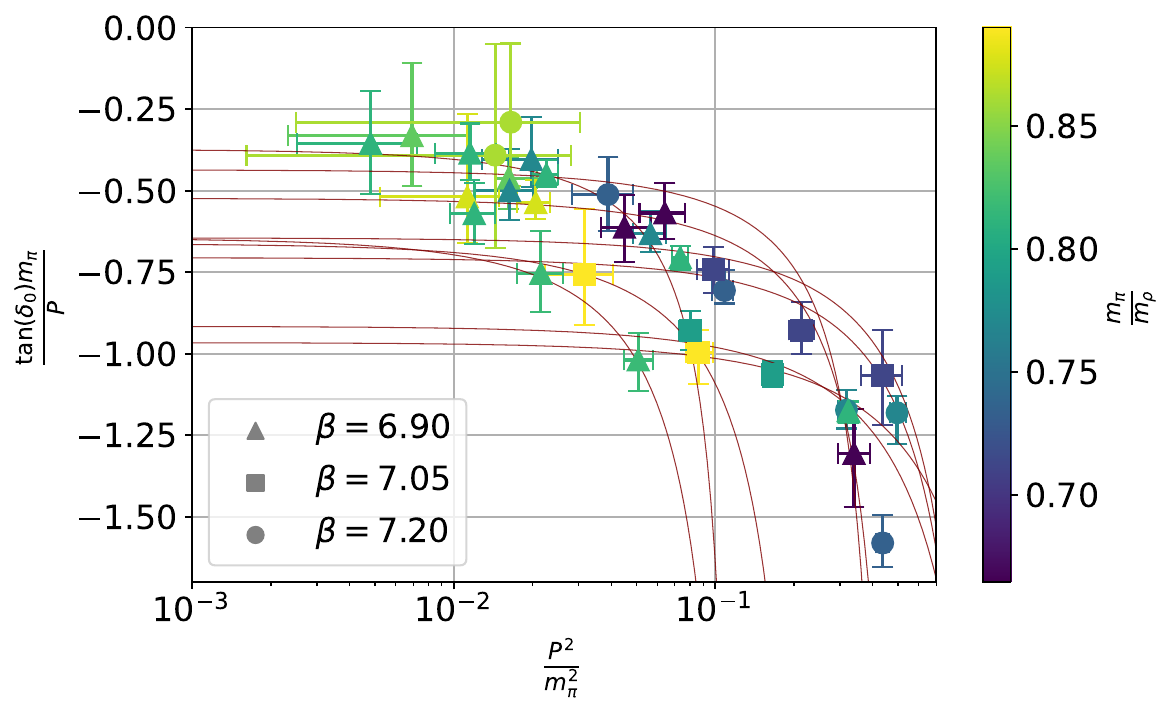}
    \caption{The dimensionless \textit{effective scattering length} plotted against the dimensionless momentum squared. The y-axis is the inverse of eq.~\eqref{eq:p_cot_PS} times the pion mass. Different gauge couplings are indicated by different symbols. The colour-coding shows the corresponding value of $\frac{m_\pi}{m_\rho}$. The thin lines show the fits that were performed to obtain the scattering length. All points and the fits are consistent with a negative scattering length.}
    \label{plot:tan_PS}
\end{center}
\end{figure}
\begin{figure}
\begin{center}
    \includegraphics[width=0.74\textwidth]{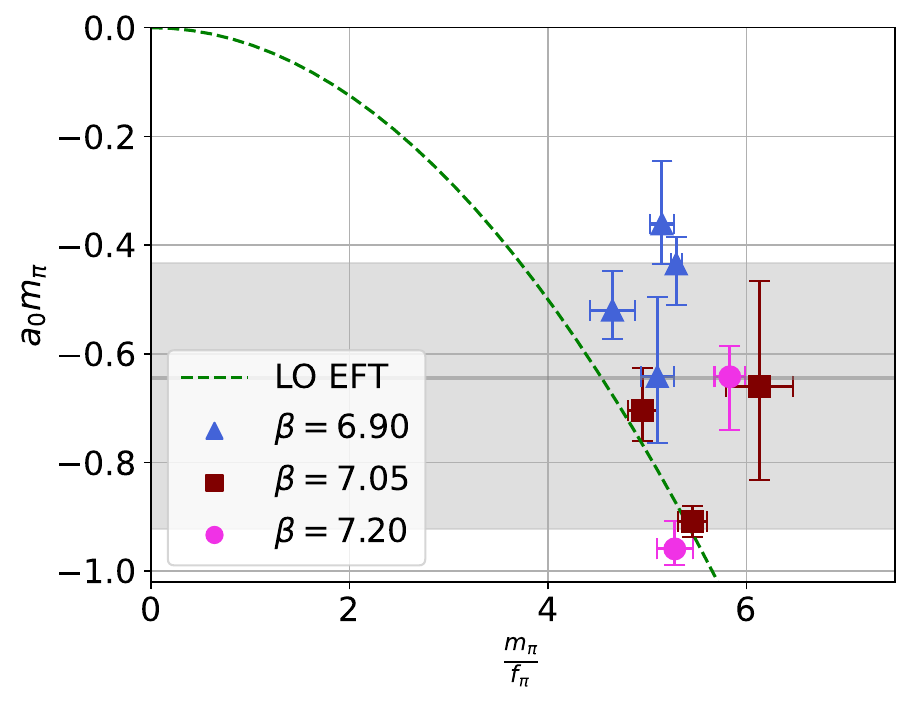}
    \caption{Results from fitting the phase shift with eq.~\ref{eq:p_cot_PS}. Shown is the scattering length $a_0m_\pi$ against the ratio of the mass to the decay constant of the pion. Different colors and symbols correspond to different values for the inverse coupling $\beta$. We observe a consistent negative scattering length across all ensembles. The horizontal grey line and band indicate the estimated central value and error for the scattering length. The green dashed line shows the expected result from leading order chiral perturbation theory.}
    \label{plot:scattering_length_fpi}
\end{center}
\end{figure}
Here, we only consider the phase shift in the isospin-2 channel. It contributes the most to the total cross-section since 14 out of the total 25 combinations of pions scatter in this channel. Furthermore, as we are interested in cosmological scattering at low relative momenta, we expect predominantly scattering in the s-wave. The isospin-2 channel with vanishing center-of-mass momentum probes s-wave scattering only. The isospin-1 channel does not probe any s-wave $2\to2$ scattering of pions, as it vanishes for zero center-of-mass momentum. We will address the other isospin channels and the $3\to2$ channel in future works \cite{Dengler:2024dm}. Also, we scale all quantities with the pion mass to give dimensionless results independent of the lattice constant. This has the distinct advantage of eliminating parts of the systematic discretization errors.

First, we take a look at the scattering phase shift obtained with eq.~\eqref{eq:tan_PS}. The phase shift contains the complete scattering information in this channel. This includes resonances which are indicated by zero-crossings. We do not find an indication for a bound state below the elastic threshold in our energy levels \cite{Dengler:2024dm}. Fig. \ref{plot:tan_PS} shows the dimensionless \textit{effective scattering length} which is given by the inverse of eq.~\eqref{eq:p_cot_PS} and coincides with $a_0m_\pi$ for $p^*\to0$. We show the leading order fits in $p^{*2}$ of eq.~\eqref{eq:p_cot_PS} as grey lines.

The resulting scattering length is shown in fig.
\ref{plot:scattering_length_fpi}. We find a consistently negative scattering length across all ensembles. The grey band shows the estimated central value of the scattering length with errors assuming a negligible mass dependence. Additionally, we depict the prediction from leading-order chiral perturbation theory. It is given by \cite{Bijnens:2011fm}
\begin{align} \label{eq:LO_chiPT}
    a_0 m_\pi = - \frac{1}{32} \left(\frac{m_\pi}{f_\pi}\right)^2.
\end{align}
For the fermion masses considered here, the order of magnitude matches the one of the chiral perturbation theory prediction, although the pion masses appear at the edge or beyond the validity of leading order chiral perturbation theory \cite{Kulkarni:2022bvh,Bennett:2017kga}. However, some data points differ from chiral perturbation theory more than two sigma. That might be an artefact of the limited number of data points used in the extrapolation towards zero momentum. At the current point, we are unable to discriminate between the behaviour predicted by chiral perturbation theory and any deviation from the predicted behaviour. Our estimated central value for the scattering length in this fermion mass range is $a_0m_\pi = -0.65^{+0.2}_{-0.3}$. We can use that to approximate the cross-section to compare it to astronomical data by $\sigma = \pi a_0^2$. We take the limit obtained in ref. \cite{Eckert:2022qia}, which constrains $\sigma/\text{m}<0.19\text{cm}^2/\text{g}$. From this we can constrain the lattice constant and with that the dark matter particle mass to $m_{DM}$>115 MeV. This is compatible with current constraints on strongly interacting dark matter \cite{Kulkarni:2022bvh} as well as earlier investigations using a more simplified analysis \cite{Zierler:2022qfq}.

\section{Conclusions}

Strongly-interacting models are a promising candidate for particle dark matter as they give solutions to known problems related to dark matter. In this paper, we used lattice field theory to study scattering properties of a specific realisation of SIMP DM. Together with studies on the mass spectrum \cite{Kulkarni:2022bvh,Bennett:2017kga,Bennett:2023rsl,Bennett:2019jzz}, these non-perturbative results can be used to determine low energy constants in an effective description with chiral perturbation theory \cite{Kulkarni:2022bvh,Bennett:2017kga}. The natural next step is the determination of the phase shift in the isospin-1 channel including the $3\to2$ process as well as further derived quantities \cite{Dengler:2024dm}.

\acknowledgments
YD and FZ have been supported the Austrian Science Fund research teams grant STRONG-DM (FG1). FZ has been supported by  the STFC Grant No. ST/X000648/1. The computations have been performed on the Vienna Scientific Cluster (VSC4).

{\bf Open Access Statement - } For the purpose of open access, the authors have applied a Creative Commons 
Attribution (CC BY) licence  to any Author Accepted Manuscript version arising.

{\bf Research Data Access Statement}---These results are preliminary. Further analysis and the data generated for this manuscript will be released together with an upcoming publication.

\bibliography{bibliography}

\begin{thebibliography}{21}
\providecommand{\natexlab}[1]{#1}
\providecommand{\url}[1]{\texttt{#1}}
\expandafter\ifx\csname urlstyle\endcsname\relax
  \providecommand{\doi}[1]{doi: #1}\else
  \providecommand{\doi}{doi: \begingroup \urlstyle{rm}\Url}\fi

\bibitem[Rubin et~al.(1978)Rubin, Ford, and Thonnard]{Rubin:1978kmz}
Vera~C. Rubin, W.~Kent Ford, Jr., and Norbert Thonnard.
\newblock {Extended rotation curves of high-luminosity spiral galaxies. IV.
  Systematic dynamical properties, Sa through Sc}.
\newblock \emph{Astrophys. J. Lett.}, 225:\penalty0 L107--L111, 1978.
\newblock \doi{10.1086/182804}.

\bibitem[Tulin and Yu(2018)]{Tulin:2017ara}
Sean Tulin and Hai-Bo Yu.
\newblock {Dark Matter Self-interactions and Small Scale Structure}.
\newblock \emph{Phys. Rept.}, 730:\penalty0 1--57, 2018.
\newblock \doi{10.1016/j.physrep.2017.11.004}.

\bibitem[Hochberg et~al.(2015)Hochberg, Kuflik, Murayama, Volansky, and
  Wacker]{Hochberg:2014kqa}
Yonit Hochberg, Eric Kuflik, Hitoshi Murayama, Tomer Volansky, and Jay~G.
  Wacker.
\newblock {Model for Thermal Relic Dark Matter of Strongly Interacting Massive
  Particles}.
\newblock \emph{Phys. Rev. Lett.}, 115\penalty0 (2):\penalty0 021301, 2015.
\newblock \doi{10.1103/PhysRevLett.115.021301}.

\bibitem[Kulkarni et~al.(2023)Kulkarni, Maas, Mee, Nikolic, Pradler, and
  Zierler]{Kulkarni:2022bvh}
Suchita Kulkarni, Axel Maas, Se\'an Mee, Marco Nikolic, Josef Pradler, and
  Fabian Zierler.
\newblock {Low-energy effective description of dark $Sp(4)$ theories}.
\newblock \emph{SciPost Phys.}, 14:\penalty0 044, 2023.
\newblock \doi{10.21468/SciPostPhys.14.3.044}.

\bibitem[Eckert et~al.(2022)Eckert, Ettori, Robertson, Massey, Pointecouteau,
  Harvey, and McCarthy]{Eckert:2022qia}
D.~Eckert, S.~Ettori, A.~Robertson, R.~Massey, E.~Pointecouteau, D.~Harvey, and
  I.~G. McCarthy.
\newblock {Constraints on dark matter self-interaction from the internal
  density profiles of X-COP galaxy clusters}.
\newblock \emph{Astron. Astrophys.}, 666:\penalty0 A41, 2022.
\newblock \doi{10.1051/0004-6361/202243205}.

\bibitem[Bennett et~al.(2018)Bennett, Hong, Lee, Lin, Lucini, Piai, and
  Vadacchino]{Bennett:2017kga}
Ed~Bennett, Deog~Ki Hong, Jong-Wan Lee, C.~J.~David Lin, Biagio Lucini,
  Maurizio Piai, and Davide Vadacchino.
\newblock {Sp(4) gauge theory on the lattice: towards SU(4)/Sp(4) composite
  Higgs (and beyond)}.
\newblock \emph{JHEP}, 03:\penalty0 185, 2018.
\newblock \doi{10.1007/JHEP03(2018)185}.

\bibitem[Del~Debbio et~al.(2010)Del~Debbio, Patella, and
  Pica]{DelDebbio:2008zf}
Luigi Del~Debbio, Agostino Patella, and Claudio Pica.
\newblock {Higher representations on the lattice: Numerical simulations. SU(2)
  with adjoint fermions}.
\newblock \emph{Phys. Rev. D}, 81:\penalty0 094503, 2010.
\newblock \doi{10.1103/PhysRevD.81.094503}.

\bibitem[Drach et~al.(2022)Drach, Fritzsch, Rago, and
  Romero-L\'opez]{Drach:2021uhl}
Vincent Drach, Patrick Fritzsch, Antonio Rago, and Fernando Romero-L\'opez.
\newblock {Singlet channel scattering in a composite Higgs model on the
  lattice}.
\newblock \emph{Eur. Phys. J. C}, 82\penalty0 (1):\penalty0 47, 2022.
\newblock \doi{10.1140/epjc/s10052-021-09914-y}.

\bibitem[Blum et~al.(2023)]{RBC:2023xqv}
Thomas Blum et~al.
\newblock {Isospin 0 and 2 two-pion scattering at physical pion mass using
  all-to-all propagators with periodic boundary conditions in lattice QCD}.
\newblock \emph{Phys. Rev. D}, 107\penalty0 (9):\penalty0 094512, 2023.
\newblock \doi{10.1103/PhysRevD.107.094512}.

\bibitem[Janowski et~al.(2019)Janowski, Drach, and Prelovsek]{Janowski:2019svg}
Tadeusz Janowski, Vincent Drach, and Sasa Prelovsek.
\newblock {Resonance Study of SU(2) Model with 2 Fundamental Flavours of
  Fermions}.
\newblock \emph{PoS}, LATTICE2019:\penalty0 123, 2019.
\newblock \doi{10.22323/1.363.0123}.

\bibitem[Foley et~al.(2005)Foley, Jimmy~Juge, O'Cais, Peardon, Ryan, and
  Skullerud]{Foley:2005ac}
Justin Foley, K.~Jimmy~Juge, Alan O'Cais, Mike Peardon, Sinead~M. Ryan, and
  Jon-Ivar Skullerud.
\newblock {Practical all-to-all propagators for lattice QCD}.
\newblock \emph{Comput. Phys. Commun.}, 172:\penalty0 145--162, 2005.
\newblock \doi{10.1016/j.cpc.2005.06.008}.

\bibitem[Umeda(2007)]{Umeda:2007hy}
Takashi Umeda.
\newblock {A Constant contribution in meson correlators at finite temperature}.
\newblock \emph{Phys. Rev. D}, 75:\penalty0 094502, 2007.
\newblock \doi{10.1103/PhysRevD.75.094502}.

\bibitem[Lepage(2021)]{peter_lepage_2021_5733391}
Peter Lepage.
\newblock gplepage/corrfitter: corrfitter version 8.2, November 2021.
\newblock URL \url{https://doi.org/10.5281/zenodo.5733391}.

\bibitem[Rummukainen and Gottlieb(1995)]{Rummukainen:1995vs}
K.~Rummukainen and Steven~A. Gottlieb.
\newblock {Resonance scattering phase shifts on a nonrest frame lattice}.
\newblock \emph{Nucl. Phys. B}, 450:\penalty0 397--436, 1995.
\newblock \doi{10.1016/0550-3213(95)00313-H}.

\bibitem[Jenny et~al.(2022)Jenny, Maas, and Riederer]{Jenny:2022atm}
Patrick Jenny, Axel Maas, and Bernd Riederer.
\newblock {Vector boson scattering from the lattice}.
\newblock \emph{Phys. Rev. D}, 105\penalty0 (11):\penalty0 114513, 2022.
\newblock \doi{10.1103/PhysRevD.105.114513}.

\bibitem[Mei\ss{}ner and Rusetsky(2022)]{Meissner:2022cbi}
Ulf-G Mei\ss{}ner and Akaki Rusetsky.
\newblock \emph{{Effective Field Theories}}.
\newblock Cambridge University Press, 8 2022.
\newblock ISBN 978-1-108-68903-8.
\newblock \doi{10.1017/9781108689038}.

\bibitem[Dengler et~al.(2024)Dengler, Maas, and Zieler]{Dengler:2024dm}
Yannick Dengler, Axel Maas, and Fabian Zieler, 2024.
\newblock (in preparation).

\bibitem[Bijnens and Lu(2011)]{Bijnens:2011fm}
Johan Bijnens and Jie Lu.
\newblock {Meson-meson Scattering in QCD-like Theories}.
\newblock \emph{JHEP}, 03:\penalty0 028, 2011.
\newblock \doi{10.1007/JHEP03(2011)028}.

\bibitem[Zierler et~al.(2023)Zierler, Lee, Maas, and Pressler]{Zierler:2022qfq}
Fabian Zierler, Jong-Wan Lee, Axel Maas, and Felix Pressler.
\newblock {Singlet Mesons in Dark $Sp(4)$ Theories}.
\newblock \emph{PoS}, LATTICE2022:\penalty0 225, 2023.
\newblock \doi{10.22323/1.430.0225}.

\bibitem[Bennett et~al.(2023)Bennett, Hsiao, Lee, Lucini, Maas, Piai, and
  Zierler]{Bennett:2023rsl}
Ed~Bennett, Ho~Hsiao, Jong-Wan Lee, Biagio Lucini, Axel Maas, Maurizio Piai,
  and Fabian Zierler.
\newblock {Singlets in gauge theories with fundamental matter}.
\newblock \emph{pre-print}, 4 2023.

\bibitem[Bennett et~al.(2019)Bennett, Hong, Lee, Lin, Lucini, Piai, and
  Vadacchino]{Bennett:2019jzz}
Ed~Bennett, Deog~Ki Hong, Jong-Wan Lee, C.~J.~David Lin, Biagio Lucini,
  Maurizio Piai, and Davide Vadacchino.
\newblock {Sp(4) gauge theories on the lattice: $N_f=2$ dynamical fundamental
  fermions}.
\newblock \emph{JHEP}, 12:\penalty0 053, 2019.
\newblock \doi{10.1007/JHEP12(2019)053}.

\end{thebibliography}

\end{document}